\documentclass[11pt,twoside]{article}

\usepackage{asp2006}
\usepackage{epsf}
\usepackage{psfig}
\usepackage{lscape}

\begin{document}
\title{The blazar's divide and the properties of Fermi blazars}   
\author{Gabriele Ghisellini}   
\affil{INAF -- Osservatorio Astronomico di Brera}    

\begin{abstract} 
The LAT instrument, onboard the {\it Fermi} satellite, 
in its first three months of operation
detected more than 100 blazars at more than the 10$\sigma$ level.
This is already a great improvement with respect to its
predecessor, the instrument EGRET onboard the {\it Compton Gamma Ray Observatory}. 
Observationally, the new detections follow and confirm the so--called blazar sequence,
relating the bolometric observed non--thermal luminosity to the overall shape of
the spectral energy distribution.
We have studied the general physical properties of all these bright {\it Fermi}
blazars, and found that their jets are matter dominated, carrying a large total 
power that correlates with the luminosity of their accretion disks.
We suggest that the division of blazars into the two subclasses of broad line emitting
objects (Flat Spectrum Radio Quasars) and line--less BL Lacs is a consequence of
a rather drastic change of the accretion mode, becoming radiatively inefficient
below a critical value of the accretion rate, corresponding to a disk
luminosity of $\sim$1 per cent of the Eddington one.
The reduction of the ionizing photons below this limit implies that the
broad line clouds, even if present, cannot produce significant broad lines,
and the object becomes a BL Lac.
\end{abstract}



\section{The Fermi blazar sequence}

The  Large Area Telescope (LAT) on board the {\it Fermi Gamma Ray Space 
Telescope (Fermi)} revealed in the first three months of operation
57 flat spectrum radio quasars (FSRQs), 42 BL Lac objects, and 5 blazars 
with uncertain classification (Abdo et al. 2009a, hereafter A09; Foschini et al.,
these proceedings).

\begin{figure}
\vskip -0.5 cm
\hskip -0.6 cm
\begin{tabular}{ll}
\psfig{figure=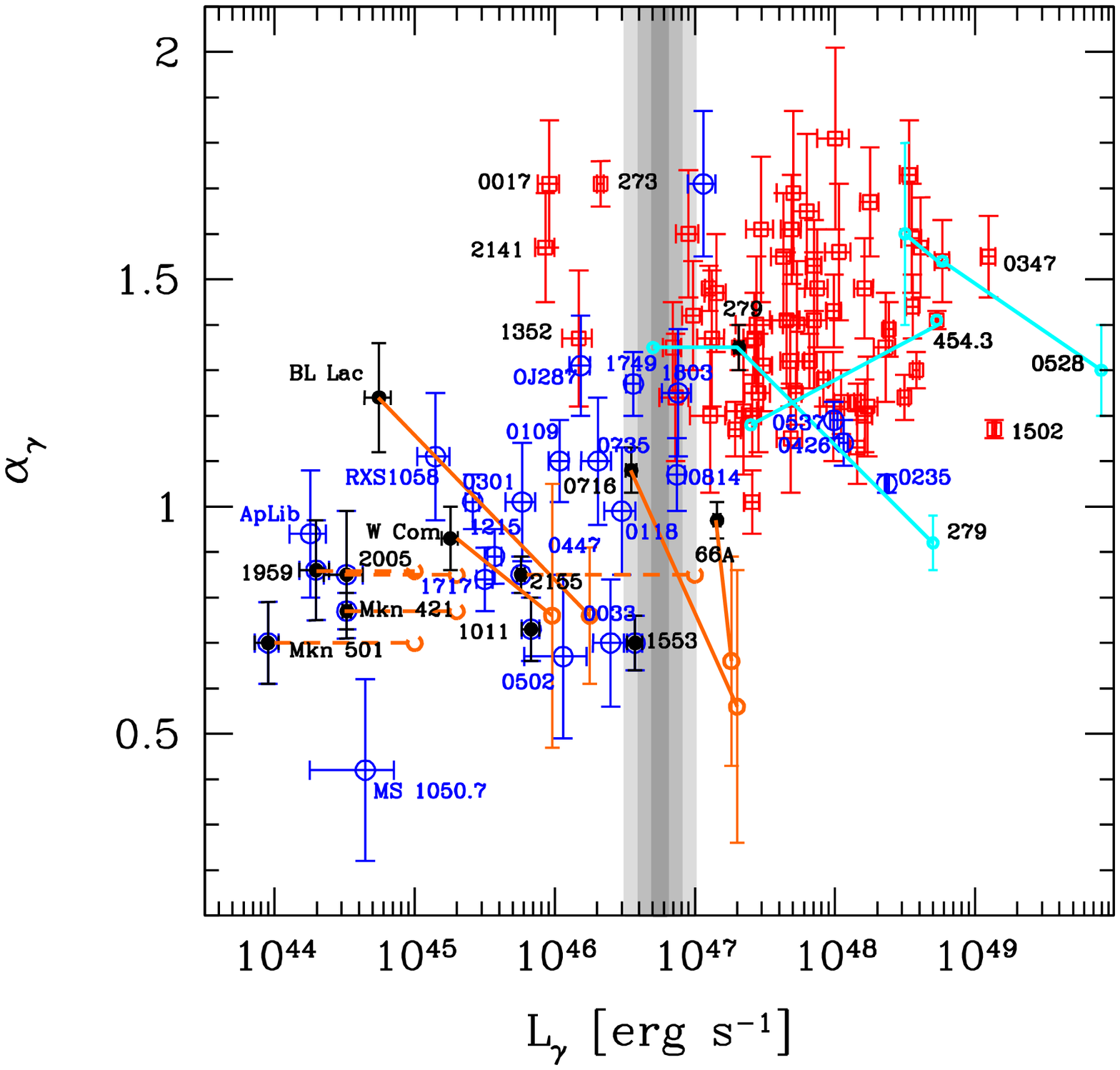,width=6.8cm,height=6.8 cm}  
&\psfig{figure=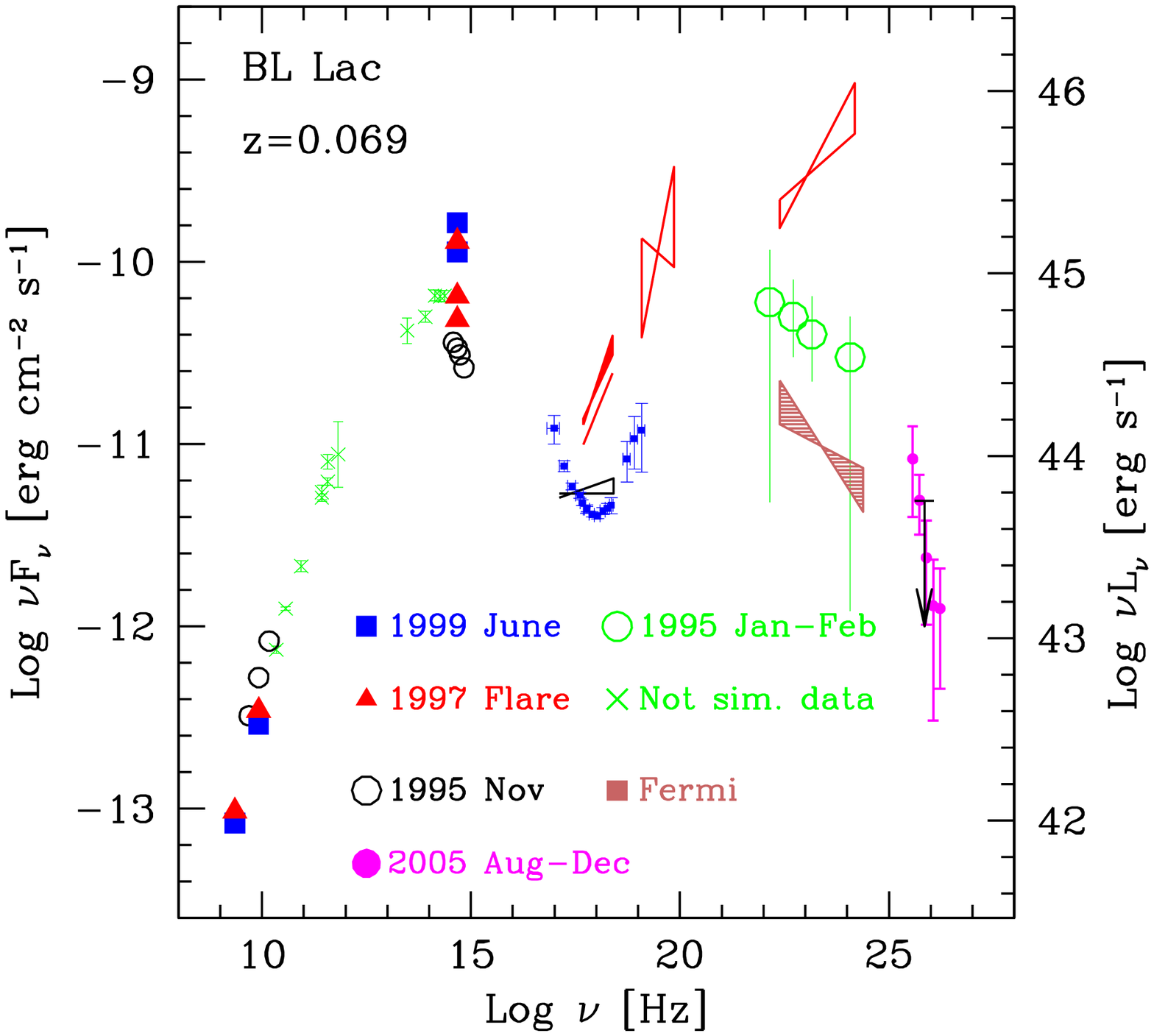,width=6.8cm,height=6.8 cm}
\end{tabular}
\vskip -0.5 cm
\caption{
{\it Left panel:}
energy spectral index vs $\gamma$--ray luminosity
for all blazars in the list of A09.
Empty squares and cirles are BL Lacs and FSRQs, respectively.
Filled symbols correspond to sources already detected in the TeV band.
For a few blazars
we show the observed range of $\gamma$--ray luminosity
and spectral index, using past EGRET or {\it AGILE} observations.
This is indicated by a segment. 
The grey stripes at about $L_\gamma=10^{47}$ erg s$^{-1}$ 
mark the divide between BL Lac objects and FSRQs.
{\it Right panel:} the SEDs of BL Lac itself illustrates the
dramatic variability of blazars, especially at high energies.
}
\label{alphal}
\end{figure}

Ghisellini et al. (2009a) showed that the spectral index $\alpha_\gamma$
correlates with the $\gamma$--ray luminosity $L_\gamma$ 
and that BL Lacs and FSRQs occupy different regions of the 
$\alpha_\gamma-L_\gamma$ plane.
There is a rather well defined boundary between BL Lacs and FSRQs as 
shown in Fig. \ref{alphal}.
Empty circles and squares correspond to BL Lac objects and FSRQs, 
respectively, while filled symbols indicate sources also detected in 
the TeV band.
This correlation holds {\it despite} the large amplitude variability
of blazars, especially at high energies.
Examples of how variability can change the position of single sources in 
the $\alpha_\gamma$--$L_\gamma$ plane are shown in Fig. \ref{alphal} by the segments
connecting the locations of specific sources at different times.
Note that several sources ``move" orthogonally to the 
correlation defined by the ensemble of sources, i.e.
they become harder when brighter (with the exception of 3C 454.3).
The high and the low $\gamma$--ray states of single
sources can be dramatically different, and this
implies that the distribution in luminosity 
within each blazar class is largely affected by
the variability of the sources.

The exceptional case of BL Lac itself is shown in the right panel of
Fig. \ref{alphal}.
Its $\gamma$--ray luminosity varied by two orders of magnitude.
Moreover the slope of the high energy emission varied 
from $\alpha_\gamma\sim 0.7$ (peak above 10 GeV) during the 1997 flare
(Bloom et al. 1997), 
to the {\it Fermi}--observed value of $\alpha_\gamma\sim 1.2$ 
(peak around or below 100 MeV), 
corresponding to the lowest observed $\gamma$--ray state.

Fig. \ref{alphal}  shows that BL Lacs and FSRQs separate at
$L_\gamma\sim 10^{47}$ erg s$^{-1}$, as indicated by the grey stripes.
Furthermore there is a less clear-cut separation 
in spectral indices, occurring at $\alpha_\gamma\simeq 1.2$.

This behavior is just what the ``blazar sequence" 
(Fossati et al. 1998; Ghisellini et al. 1998)
would predict: low power BL Lac objects peak at higher energies,
with the high energy peak often located beyond the LAT range:
they have smaller $L_\gamma$ and flatter $\alpha_\gamma$.
FSRQs, instead, peak at lower frequencies, and the peak of their
high energy emission (dominating their power output) is below 100 MeV.
In the LAT energy range they are steep, but powerful.
Therefore the left panel of Fig. \ref{alphal} represents the $\gamma$--ray 
selected version of the blazar sequence.

\subsection{The divide}

The other intriguing feature of Fig. \ref{alphal}
is the existence of a $\gamma$--ray luminosity
dividing BL Lacs from FSRQs.
We have proposed that this is a consequence of the change
of the accretion regime, becoming radiatively inefficient
below a critical disk luminosity, in units of Eddington. 
This reflects also in a critical (dividing) luminosity of the 
observed beamed emission, rather well tracked by $L_\gamma$.
To understand why in a simple way, assume
that most of the bright blazars detected by the 3--months LAT survey 
have approximately the same black hole mass.
Assume also that the largest $L_\gamma$ correspond to jets 
with the largest power carried in bulk motion of particles and fields.
Finally, assume that the jet power and the accretion rate are related.
These three assumptions, that will be better justified later,
imply that the most luminous blazars have the most powerful jets and
are accreting near Eddington. 
These are the FSRQs with $L_\gamma\sim 10^{49}$ erg s$^{-1}$.
The dividing $L_\gamma$ is a factor 100 less, so it should correspond
to disks emitting at the 1\% level of the Eddington level.
Below this value we find BL Lacs, that have no (or very weak) 
broad emission lines. 
If the disk becomes radiatively inefficient at $L_{\rm d} < 10^{-2}L_{\rm Edd}$
the broad line region receives a much decreased ionizing luminosity,
and the lines become much weaker.
The radiation energy density of the lines becomes unimportant for 
the formation of the high energy continuum (there are much less 
seed photons for the Inverse Compton process), implying: 
i) a reduced ``Compton dominance" (i.e. the ratio of the Compton to synchrotron
luminosities);
ii) less severe cooling for the emitting electrons, that can then
achieve larger energies and then
iii) a shift of both the synchrotron and the Inverse Compton peak frequencies 
to larger values.

According to this interpretation, it is the accretion mode that determines
the ``look" of the radiation produced by the jet, not a property of the jet itself.

\section{General properties of the {\it Fermi} blazars}

We (Ghisellini et al. 2009b, hereafter G09) have studied and modelled almost all 
the {\it Fermi} blazars detected in its first 3--months of operation.
We excluded objects without known redshift and a few
with very few data available (insufficient to construct a 
meaningful SED).
In total, we studied 85 blazars (including one Narrow Line Seyfert 1,
see Abdo et al. 2009b, 2009c).

\begin{figure}
\vskip -0.5 cm \hskip -0.4 cm
\begin{tabular}{ll}
\psfig{figure=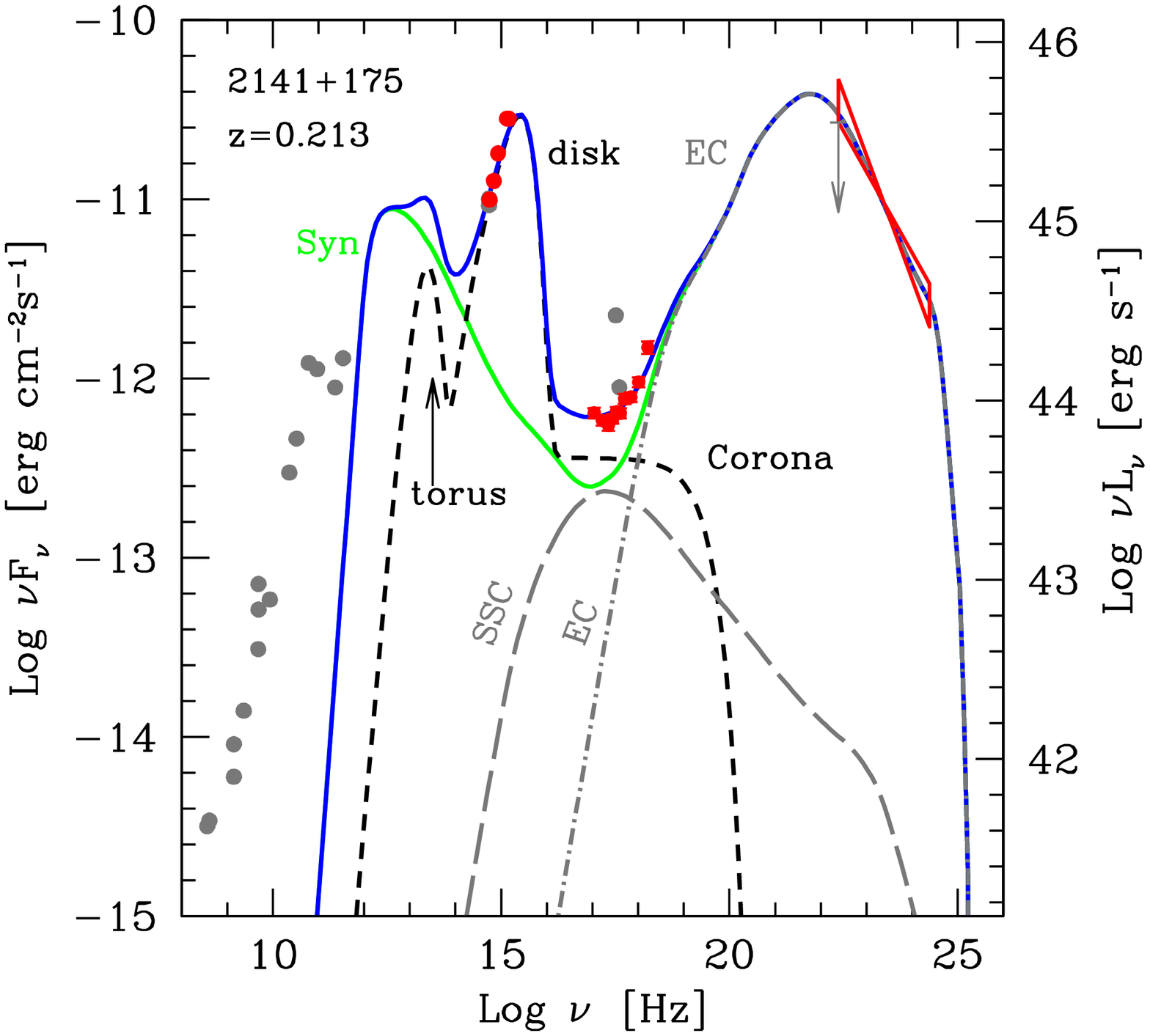,width=6.8cm,height=6.8 cm}
&\psfig{figure=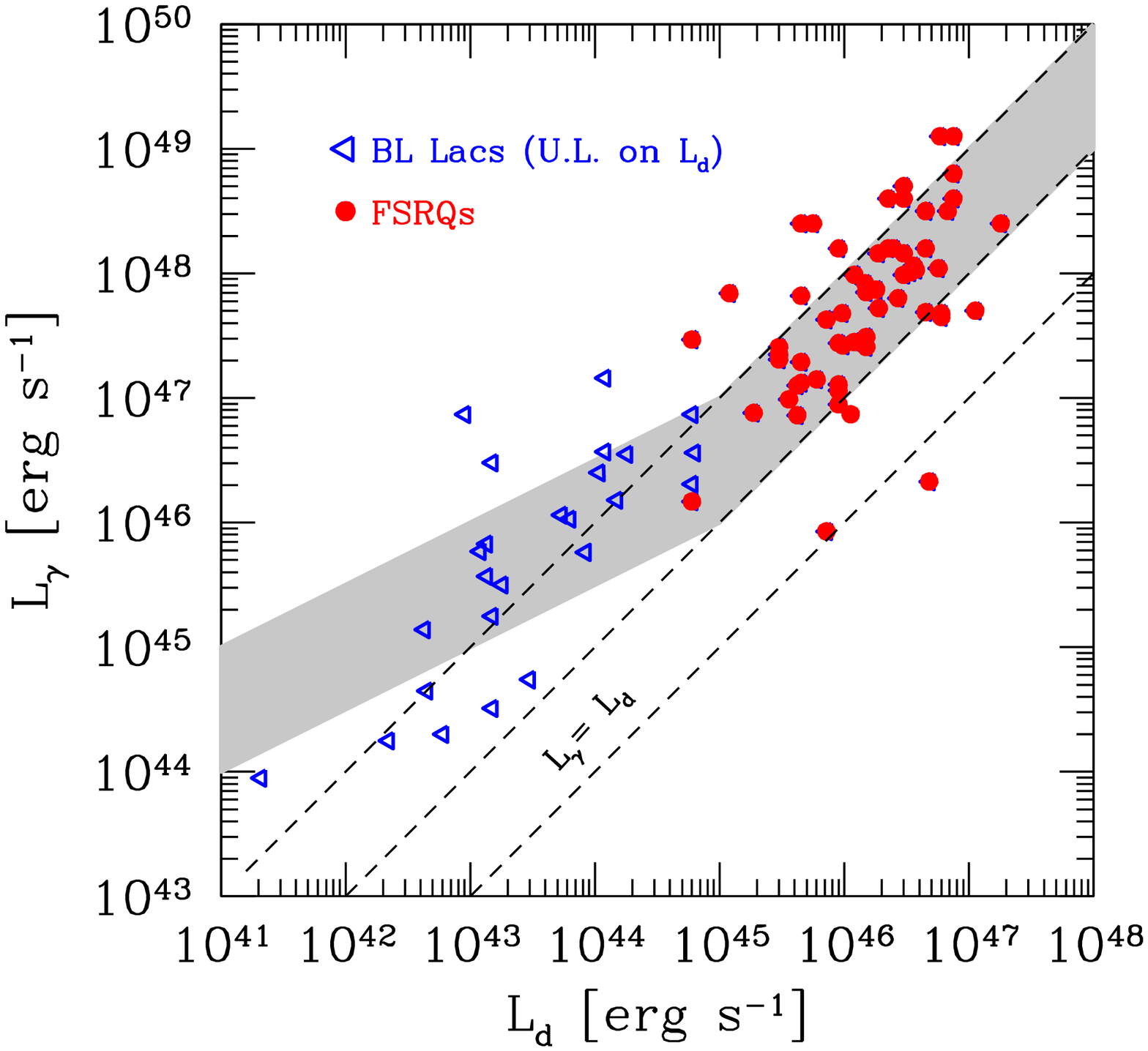,width=6.8cm,height=6.8 cm}
\end{tabular}
\vskip -0.5 cm
\caption{
{\it Left panel:} the SED of the FSRQ 2141+175 and the fitting model.
We label the different components. Note how the synchrotron spectrum,
peaking at low frequencies, makes the accretion disk flux ``naked".
In this cases the data are good enough for estimating both the
black hole mass and the accretion rate.
{\it Right panel:}
the $\gamma$--ray luminosity as a function 
of the accretion disk luminosity for {\it Fermi} blazar of the A09 sample.
Red filled circles are FSRQs, triangles are for BL Lacs with only an
upper limit for their disk luminosity.
The grey band corresponds to what expected if the FSRQs 
with $L_d>10^{45}$ erg s$^{-1}$ have standard accretion disks 
with $L_{\rm d}>10^{-2}L_{\rm Edd}$ and $L_\gamma\propto L_{\rm jet}\propto 
\dot M\propto L_{\rm d}$,
while BL Lac have ``ADAF" like accretion with $L_{\rm d}\propto \dot M^2$.
In this case their $L_\gamma \propto L_{\rm jet}\propto 
\dot M\propto L^{1/2}_{\rm d}$.
}
\label{sed}
\end{figure}

Many of them were observed by the X--ray (XRT) and UV--optical (UVOT) telescopes 
onboard {\it Swift}, and this was of great help in characterizing their SED.
What was an exception in the EGRET era (and a result of huge efforts by many people
involved in multi-wavelength campaigns) is now routine.
Fig. \ref{sed} shows, for illustration, the SED of the FSRQ 2141+175.
As can be seen, the synchrotron spectrum peaks at very low frequencies, and
the flux produced by the accretion disk is well visible.
In this case the data are good enough to fit the optical--UV flux
with a standard Shakura --Sunjaev (1973) accretion disk.
In turn, this allows to estimate the mass of the black hole and the accretion
rate.

For these FSRQs (with good optical--UV coverage) we can then study in  
a reliable way the connection between the jet power and the accretion luminosity,
also in units of the Eddington one.
A first result is shown in the right panel of Fig. \ref{sed}:
the observed $\gamma$--ray luminosity $L_\gamma$ is related with the accretion disk
luminosity $L_{\rm d}$.
Note that for BL Lacs we have only an upper limit on $L_{\rm d}$
(shown by the triangles).

The grey stripe shows a linear relation above $L_{\rm d}=10^{45}$
erg s$^{-1}$ (with scatter, blazars can vary
their non--thermal luminosity even by one or two order of magnitude).
This is appropriate for all {\it Fermi} FSRQs.
Below this critical luminosity value there are only BL Lacs,
and the grey stripe becomes $L_\gamma \propto L_{\rm d}^{1/2}$.
This corresponds to assume that the jet power (and then the observed
luminosity, for aligned sources) scales always as the accretion rate $\dot M$,
while the disk luminosity, which is linear with $\dot M$ at high
rates, scales as $L_{\rm d} \propto \dot M^2$ below $L_{\rm d}=10^{45}$
erg s$^{-1}$, so that  $L_\gamma\propto \dot M\propto L_{\rm d}^{1/2}$
(see Ghisellini \& Tavecchio 2008).
Note that, for a $10^9 M_\odot$ black hole, this ``dividing" luminosity
corresponds to $L_{\rm d}\sim10^{-2} L_{\rm Edd}$.
In the near future, when blazars with black holes of smaller masses 
will be observed, this clear--cut division will become fuzzier.

\subsection{Jet powers}

Several attempts have been done in the past to find the
jet power and the accretion disk luminosity in blazars
and radio--loud objects in general (starting from 
Rawlings \& Saunders 1991; 
Celotti et al. 1997; 
Cavaliere \& D'Elia 2002; 
Maraschi \& Tavecchio 2003; 
Padovani \& Landt 2003; 
Sambruna et al. 2006;
Allen et al. 2006;
Celotti \& Ghisellini 2008;
Ghisellini \& Tavecchio 2008; 
Kataoka et al. 2008).
These works found large jet powers, often larger
than the luminosity produced by the disk.
However, there were two caveats: the first concerns the low energy end of the 
emitting particle distribution, where most of the electrons are.
To the end of estimating the jet power, this is a crucial quantity
if one assumes that there is one proton per electron (and this assumption
is the second caveat).
But in powerful sources, for which the radiative cooling is severe,
even low energy electrons cool in a light crossing time, leaving much
less uncertainty about the presence of low energy electrons,
distributed in energy  $\propto \gamma^{-2}$.

\begin{figure}
\vskip -0.5 cm \hskip -0.7 cm
\begin{tabular}{ll}
\psfig{figure=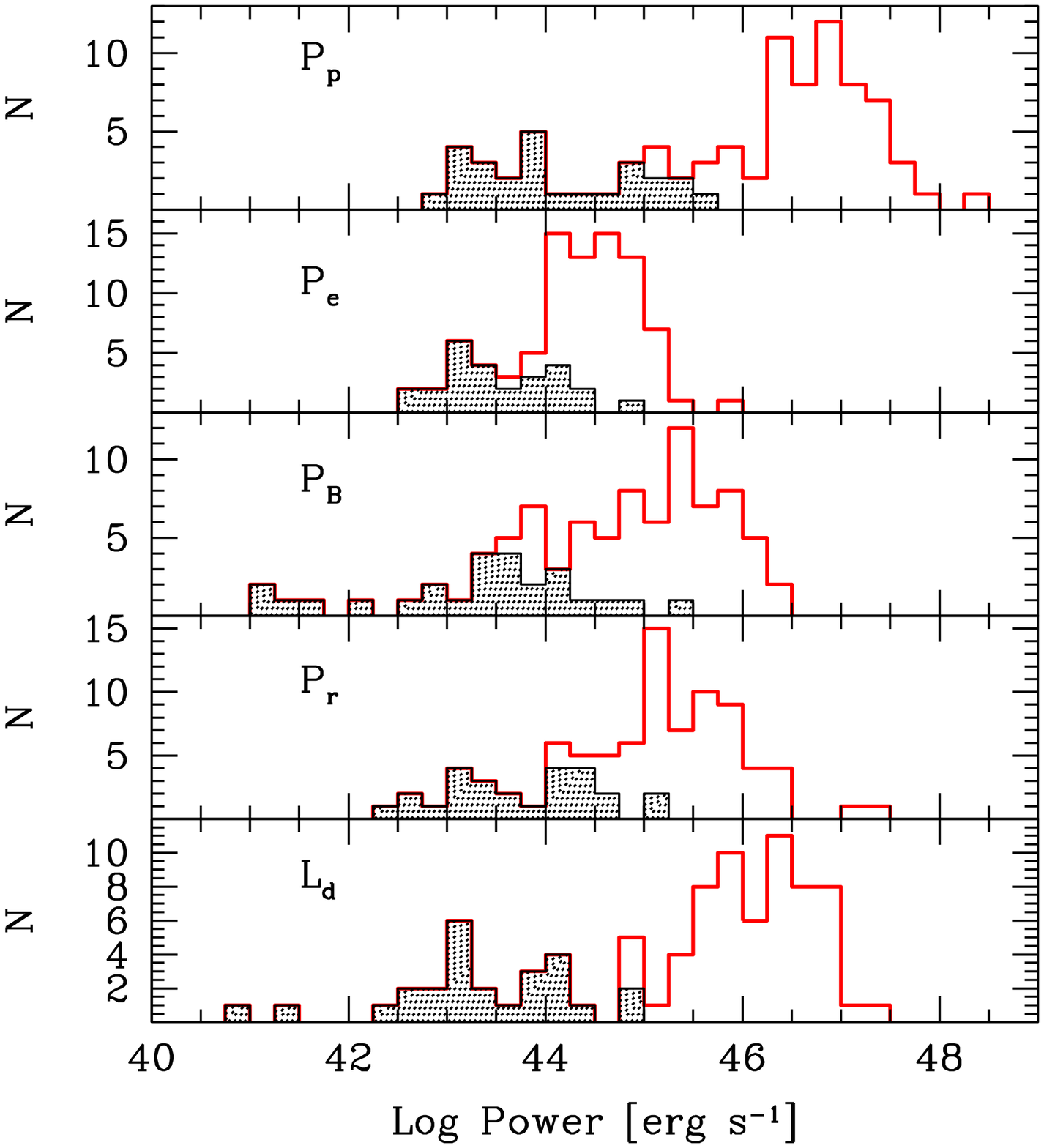,width=7cm,height=7cm}
&\psfig{figure=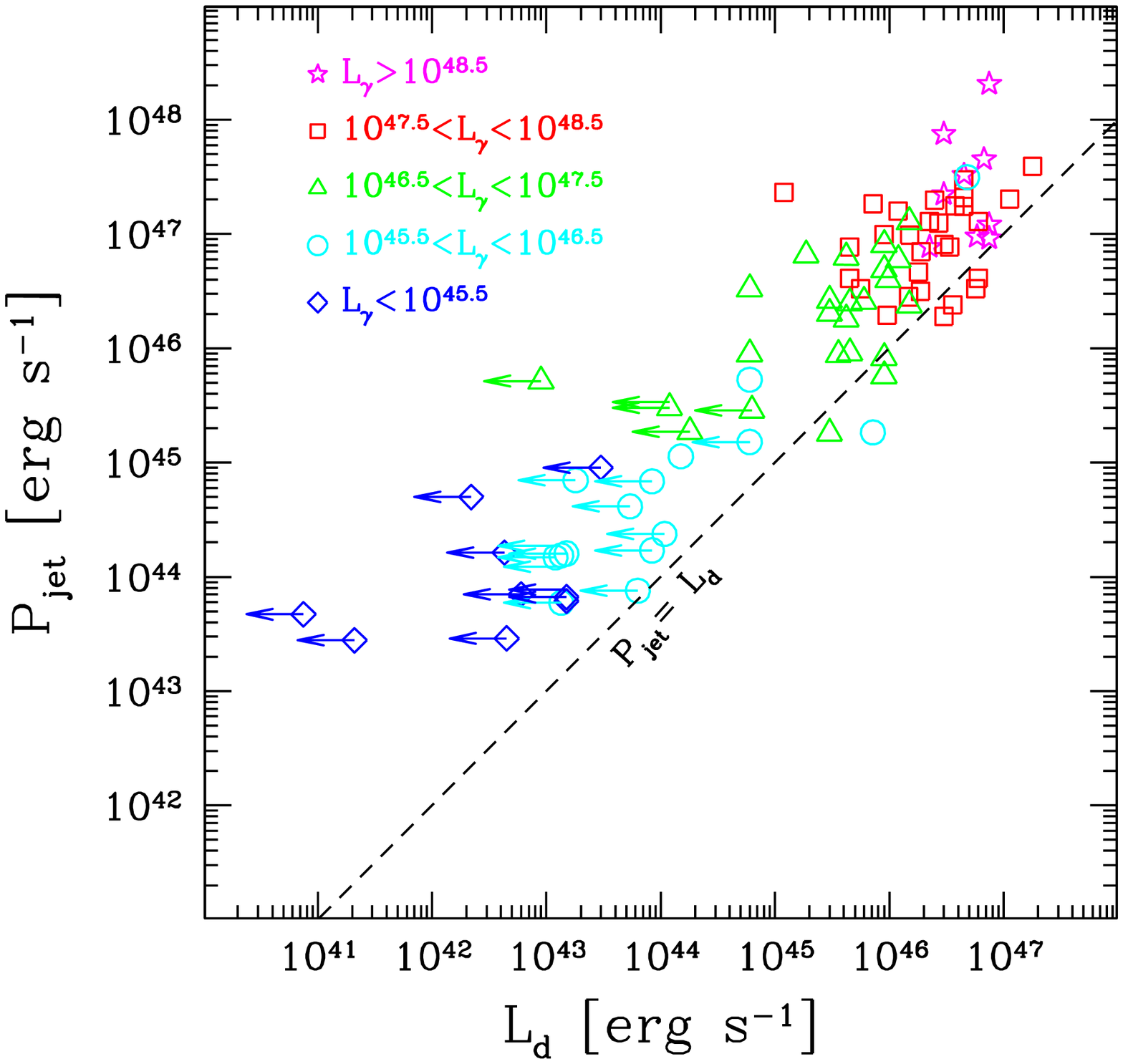,width=7cm,height=7 cm}
\end{tabular}
\vskip -0.2cm
\caption{
{\it Left panel:} The distribution of jet powers in the form of bulk motion
of cold protons ($P_{\rm p}$), emitting electrons ($P_{\rm e}$),
magnetic field ($P_{\rm B}$) and radiation ($P_{\rm r}$). 
The bottom panel shows the distribution of disk luminosities $L_{\rm d}$.
Grey shaded areas correspond to BL Lacs.
{\it Right panel:} the total jet power $P_{\rm jet}$ vs the accretion disk
luminosity $L_{\rm d}$. 
To estimate $P_{\rm jet}$, we have assumed one proton per emitting electron.
}
\label{pjld}
\end{figure}

Sikora \& Madehski (2000) and Celotti \& Ghisellini (2008) argued that electron--positron
pairs cannot be dynamically important, corresponding to a limit of a few
pairs per proton.
This issue (discussed at length in Celotti \& Ghisellini 2008 and G09) 
can be understood looking at the left panel of
Fig. \ref{pjld}, showing the histograms of the different forms of power 
carried by the jet.
The shaded areas correspond to BL Lacs.
The crucial power, that is almost model--independent, is the power $P_{\rm r}$ 
spent by the jet to produce its radiation. 
It is simply the observed, beamed, bolometric luminosity multiplied by
$\Gamma^2/\delta^4 \sim 1/\delta^2$.
For FSRQs, the distribution of $P_{\rm r}$ extends to larger values than
the distribution of $P_{\rm e}$, the power carried by the jet in the form
of emitting electrons.
So the radiation we see cannot originate by electrons (or pairs) only.
Can it come from the Poynting flux (by e.g. reconnection)?
The distribution of $P_{\rm B}$ is at slightly smaller values than the 
distribution of $P_{\rm r}$, indicating that the Poynting flux cannot
be at the origin of the radiation we see.
As described in Celotti \& Ghisellini (2008), this is a direct consequence
of the large values of the Compton dominance (i.e. the ratio of 
the Compton to the synchrotron luminosity is small), since
this limits the value of the magnetic field.

To justify the power that the jet carries in radiation we are forced to consider 
protons.
If there is one proton per electron (i.e. no pairs), then
$P_{\rm p}$ for FSRQs is a factor $\sim$10--100 larger than $P_{\rm r}$,
meaning an efficiency of 1--10\% for the jet to convert its bulk kinetic
motion into radiation.
This is reasonable: most of the jet power in FSRQs goes to form and energize the 
large radio structures, and not into radiation.
                       
We then conclude that jets should be matter dominated, at least at the 
scale (hundreds of Schwarzschild radii from the black hole) 
where most of their luminosity is produced. 
The bottom left panel of Fig. \ref{pjld} shows the distribution of the 
disk luminosities.
In this case the shaded area corresponds to upper limit for BL Lac objects,
and not to actual values.
This $L_{\rm d}$ distribution lies at intermediate values between 
$P_{\rm r}$ and $P_{\rm p}$.

\subsection{Jet powers and disk luminosities}

The right panel of Fig. \ref{pjld} shows the total jet power
$P_{\rm jet}\equiv P_{\rm p}+P_{\rm e}+P_{\rm B}$ as
a function of the thermal disk luminosity.
Arrows corresponds to BL Lacs for which only an upper limit on $L_{\rm d}$ 
could be derived.
The different symbols corresponds to blazars of different
$\gamma$--ray luminosities, and one can see that $L_\gamma$
correlates both with $P_{\rm jet}$ and $L_{\rm d}$.

As discussed in G09, there is a significant correlation between
$P_{\rm jet}$ and $L_{\rm d}$ for FSRQs, which remains highly significant
even when excluding the common redshift dependence.
The slope of this correlation is consistent with being linear,
and $P_{\rm jet}$ is larger than $L_{\rm d}$ for almost all sources,
and must be much larger for BL Lacs.

\begin{figure}
\vskip -0.5 cm
\hskip -1.3cm
\begin{tabular}{ll}
\psfig{figure=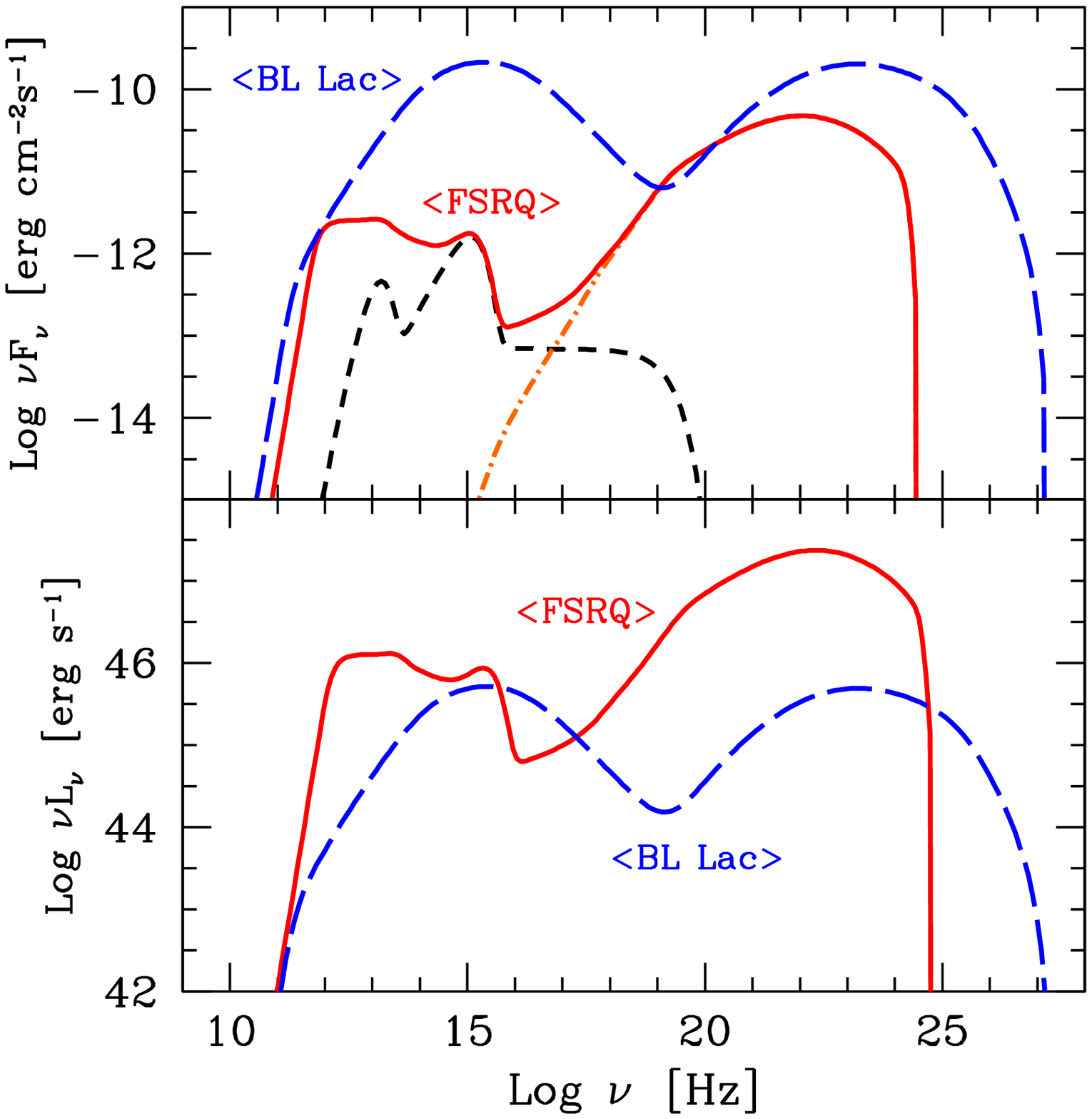,width=7.5cm,height=8 cm}
&\psfig{figure=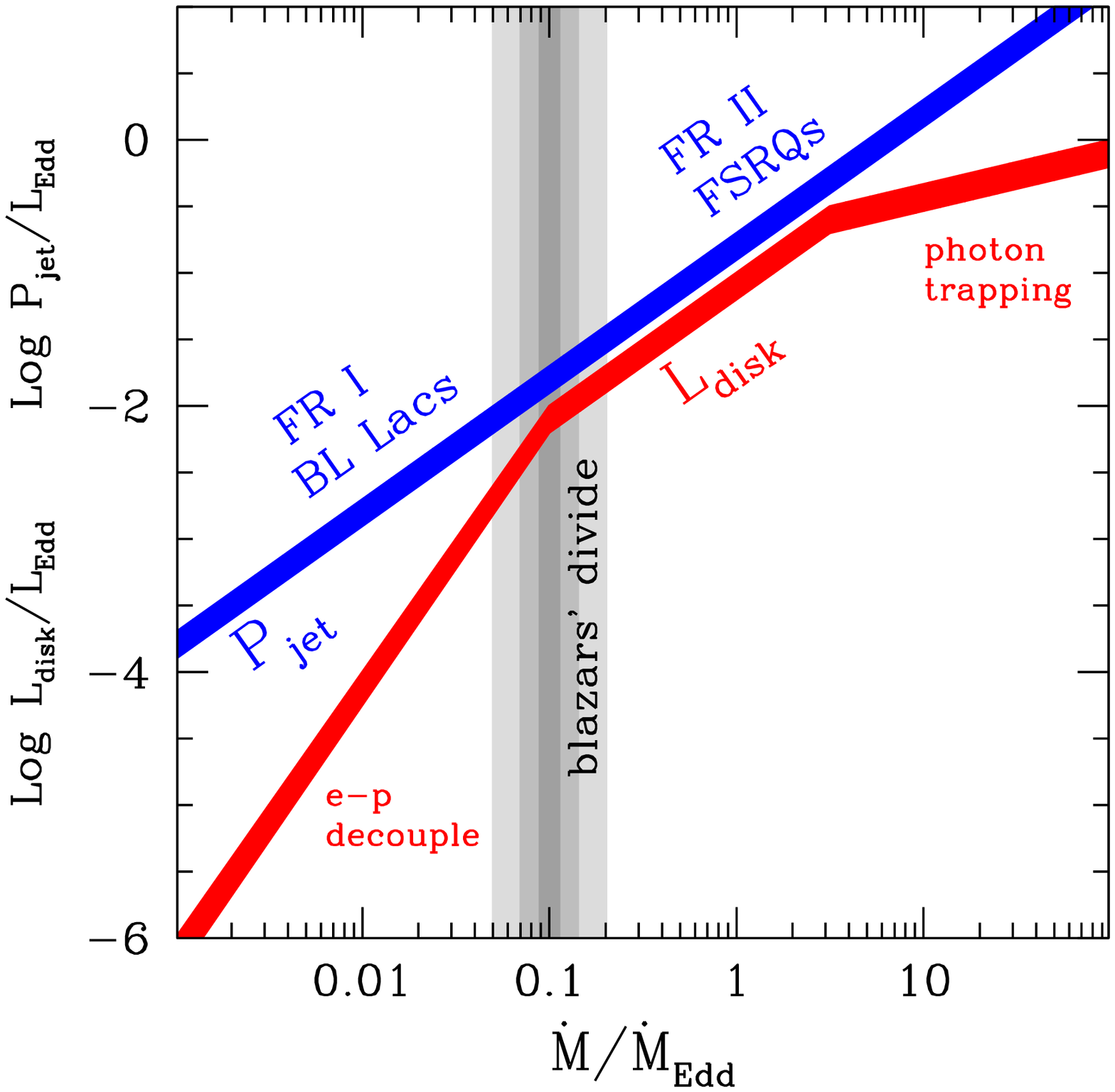,width=7.5cm,height=8 cm}
\end{tabular}
\vskip -0.5 cm
\caption{
{\it Left panel:} The average SED for BL Lacs (blue long dashed),
and FSRQs (red solid) detected in the 3--months {\it Fermi} survey,
both in $\nu F_\nu$ (top) and $\nu L_\nu$ (bottom). 
{\it Right panel:}
sketch illustrating $P_{\rm jet}$ and $L_{\rm d}$ as
a function of $\dot M/\dot M_{\rm Edd}$.
It is assumed that the jet power always scales linearly with $\dot M$,
while accretion rates below a critical value produce radiatively inefficient
accretion disks. 
In this case the object looks like a BL Lac (if aligned) or a FR I (if misaligned).
The gray stripes indicate the critical $\dot M \dot M_{\rm Edd}\sim 0.1$, 
producing the blazars' divide at $L_{\rm d}/L_{\rm Edd}\sim 10^{-2}$.
}
\label{mdot}
\end{figure}

\section{Discussion}

The first results of {\it Fermi} confirm the idea that blazars
form a sequence.
Fig. \ref{mdot} shows the average model SED constructed for BL Lacs and FSRQs
by averaging the parameters obtained by fitting the sources one by one.
It shows both the $\nu F_\nu$ and $\nu L_\nu$ representations.
In the LAT energy range the average BL Lac has a flat ($\alpha_\gamma<1$)
spectrum, while FSRQs are steeper than unity.
This is associated with the larger Compton dominance in FSRQs,
in turn associated with the presence of external seed photons for
the scattering process.
Also shown (short dashed line) is the averaged disk spectrum of FSRQs, 
together with the spectrum produced by the 
X--ray corona and the re--emission of part of the disk 
optical--UV radiation by an absorbing torus.

\subsection{Relevance of the accretion rate}

The relation between $P_{\rm jet}$ and $L_{\rm d}$ strongly suggests that 
\begin{equation}
P_{\rm jet}\, \approx \, \dot M c^2
\end{equation}
while the accretion disk luminosity 
\begin{eqnarray}
L_{\rm d}\, &\sim & \, 0.1 \dot M \, c^2 \qquad \qquad \,\,\, 
\dot M \ge \dot M_{\rm c} \nonumber \\
L_{\rm d}\, &\sim & \, 0.1 \left( { \dot M \over \dot M_{\rm c} }\right)^2 c^2 
\qquad \dot M \le \dot M_{\rm c} 
\end{eqnarray}
where the $L_{\rm d}\propto \dot M^2$ dependence is appropriate for
advection dominated accretion flows (ADAF, e.g. Narayan, Garcia \& McClintock 1997).
Radiatively inefficient disk may also correspond to 
adiabatic inflow--outflows (ADIOS, Blandford \& Begelman 1999) or a
convection dominated flows (CDAF, Narayan, Igumenshchev \& Abramowicz 2000).
At the other extreme of accretion rates (i.e. nearly Eddington) the  
density close to the hole may correspond to scattering optical depths 
larger than unity, trapping a fraction of photons and making them to be swallowed
by the black hole before escaping.
Fig. \ref{mdot} sketches the expected behavior of both $P_{\rm jet}$
and $L_{\rm d}$ as a function of $\dot M/\dot M_{\rm Edd}$, where
$\dot M_{\rm Edd} \equiv L_{\rm Edd}/c^2$.
According to this scenario all radio loud objects of all powers have
a jet power proportional to $\dot M$, irrespective of the accretion regimes.
These instead affect the emitted disk luminosity $L_{\rm d}$ at both
ends of the $\dot M$ range.
Below $L_{\rm d} \sim 10^{-2} L_{\rm Edd}$, corresponding to 
$\dot M \sim 0.1\dot M_{\rm Edd}$,
the disk becomes radiatively inefficient, its ionizing radiation is
greatly reduced, as are the broad lines. 
These objects are BL Lacs if pointing in our direction, and FR I radio--galaxies
if they point somewhere else.
Above the critical $\dot M$, jet powers and disk luminosities scale linearly,
producing a FSRQ or a powerful FR II.

\subsection{What powers blazars' jets?}

The fact that the jet power correlates with $L_{\rm d}$,
but tends to be larger than that, leads us to ask:
What is the source of the power of the jet?
Is it only the gravitational energy of the accreting matter or do we 
necessarily need also the rotational energy of a spinning black hole?
In G09 we have discussed two possible alternatives.

The first possibility stems out from the idea 
by Jolley et al. (2009) and Jolley \& Kuncic (2008), who propose that,
in jetted sources, a sizeable fraction of the accretion power
goes to power the jet.
As a result, the remaining power for the disk
luminosity is less than usually estimated
by setting $L_{\rm d}=\eta \dot M_{\rm in} c^2$, with $\eta\sim0.08$--0.1.
This implies that the mass accretion rate needed to sustain a given $L_{\rm d}$ 
is {\it larger} than what we have estimated.
Also the total accretion power is larger, and it is sufficient to
explain the derived large jet powers.

The second alternative is the more standard Blandford \& Znajek (1978)
scenario, in which jets are powered by the rotational energy of the spinning
black hole. 
In this scenario the correlation between jet power and disk luminosity
is provided by the requirement of having a sufficiently strong magnetic field,
anchored to the disk, to tap the spin energy of the hole.
If the magnetic energy density scales with the disk density,
in turn linked to the accretion rate, then $P_{\rm jet}$ should scale as $L_{\rm d}$.

\acknowledgements 
I gratefully thank my collaborators A. Celotti, L. Foschini,
G. Ghirlanda, L. Maraschi and F. Tavecchio.
I thank the grant PRIN--INAF 2007 for partial funding.


\end{document}